\documentclass[runningheads]{svmult}

\usepackage{makeidx}   
\usepackage{graphicx}  
\usepackage{subeqnar}  
\usepackage{multicol}  
\usepackage{physprbb}  
\makeindex             

\begin{document}
\title*{Ices in Star-Forming Regions: First Results from VLT-ISAAC}
\toctitle{Ices in Star-Forming Regions}
%
%
\titlerunning{Ices in Star-Forming Regions}
%
\author{Ewine F.\ van Dishoeck\inst{1,4}
\and E. Dartois\inst{2}
\and W.F.\ Thi\inst{1}
\and L. d'Hendecourt\inst{2}
\and A.G.G.M. Tielens\inst{3}
\and P.\ Ehrenfreund\inst{4}
\and W.A. Schutte\inst{4}
\and K.\ Pontoppidan\inst{1}
\and K.\ Demyk\inst{2}
\and J.\ Keane\inst{3}
\and A.C.A. Boogert\inst{5}
}
\authorrunning{E.F. van Dishoeck et al.}
%
%
\institute{Leiden Observatory, P.O. Box 9513, NL-2300 RA Leiden, The Netherlands
\and `Astrochimie Exp\'erimentale', Universit\'{e} Paris XI, Bat.\ 121,
     F-91405 Orsay, France
\and Kapteyn Institute/SRON Groningen, P.O. Box 800, NL-9700 AV Groningen,
   The Netherlands
\and Raymond \& Beverly Sackler Laboratory for Astrophysics, Leiden
Observatory, P.O. Box 9513, NL-2300 RA Leiden, The Netherlands
\and Div.\ of Physics, Mathematics \& Astronomy, Caltech, Pasadena, USA
}

\maketitle              

\begin{abstract}
The first results from a VLT-ISAAC program on the infrared spectroscopy of
deeply-embedded young stellar objects are presented.  The advent of
8-m class telescopes allows high $S/N$ spectra of low-luminosity sources to
be obtained. In our first observing run, low- and medium-resolution
spectra have been measured toward a dozen objects, mostly in the Vela
and Chamaeleon molecular clouds. The spectra show strong absorption of
H$_2$O and CO ice, as well as weak features at `3.47' and 4.62 
$\mu$m. No significant solid CH$_3$OH feature at 3.54
$\mu$m is found, indicating that the CH$_3$OH/H$_2$O ice abundance is
lower than
toward some massive protostars.  Various evolutionary diagnostics are
investigated for a set of sources in Vela.
\end{abstract}

\section{Introduction}

Interstellar matter provides the basic building blocks from which new
solar systems like our own are made. The formation of stars and
planetary systems begins with the collapse of a dense interstellar
cloud core, a reservoir of dust and gas from which the protostar and
circumstellar disk are assembled.  In this cold and dense phase,
molecules freeze-out onto the grains and form an icy mantle
surrounding the silicate and carbonaceous cores. The ices can contain
up to 40 \% of the condensible material (i.e., C and O)
(Whittet et al.\ 1998, d'Hendecourt et al.\ 1999). Much of this
material is incorporated into the circumstellar disks and ultimately
in icy solar system bodies such as comets (Ehrenfreund et al.\ 1997).
A central quest in star formation and astrochemistry is to
understand the evolution of these species from interstellar clouds to
planetary bodies, and use them as diagnostic probes of the thermal
history and physical processes (van Dishoeck \& Blake 1998,
Ehrenfreund \& Charnley 2000, Langer et al.\ 2000).

In recent years, much progress has been made in our understanding of
the chemical evolution during star formation through combined
submillimeter and infrared observations. Submillimeter
observations using telescopes such as the JCMT, CSO, IRAM 30m and SEST
probe the gas-phase composition of the warm and dense envelopes around
deeply-embedded protostars. Many different species with abundances
down to $10^{-11}$ with respect to H$_2$ can be probed through their
pure rotational transitions. Because of the high spectral resolution
of the heterodyne technique ($R=\lambda/\Delta \lambda>10^6$), the
line profiles are resolved and provide information on the location of
the molecules (e.g., quiescent gas vs.\ outflow). Since the lines are
in emission, a map of their distribution can be made.

Infrared spectroscopy provides complementary information: at these
wavelengths the vibrational modes of both gas-phase and solid-state
species can be observed, but only down to abundances of $\sim 10^{-7}$
with respect to H$_2$ at typical resolving powers $R\approx$ a few
thousand. Symmetric molecules, such as H$_3^+$, CH$_4$, C$_2$H$_2$ and
CO$_2$ have no dipole-allowed rotational transitions, and can
therefore only be probed through their strong infrared vibrational
transitions.  Moreover, PAH emission features appear throughout the
mid-infrared range, and the dominant interstellar molecule, H$_2$, has
its pure rotational transitions at these wavelengths.  The cold gases
and ices are usually observed in absorption toward an embedded YSO,
where the hot dust in the immediate circumstellar environment provides
the continuum. This technique samples only a pencil beam line of
sight.  The {\it Short Wavelength Spectrometer} (SWS) on the {\it
Infrared Space Observatory} (ISO) has opened up the mid-infrared
wavelength region and has obtained spectra of more than a dozen
protostars without atmospheric interference (see van Dishoeck \&
Tielens 2001 for review).  However, ISO was limited to the most
luminous, massive YSOs ($L\approx 10^4-10^5$ L$_{\odot}$). A major
goal of our VLT-ISAAC program is to extend this work to lower-mass
objects ($L\leq 10^3$ L$_{\odot}$) representative of our proto-Sun.

Together, the submillimeter and infrared data have led to the
following scenario for high-mass objects. In the cold pre-stellar
cores and collapsing envelopes, gas-phase molecules freeze-out onto
the grains and form an icy mantle. Here the abundances can be further
modified by grain surface reactions and, perhaps, photoprocessing of
ices. In particular, the hydrogenation and oxidation of accreted C, O,
N and CO can lead to CH$_4$, H$_2$O, NH$_3$, H$_2$CO, CH$_3$OH and
CO$_2$, respectively (Tielens \& Charnley 1997).  Most of these
species have been firmly identified in interstellar ices (e.g.,
Whittet et al.\ 1996, d'Hendecourt et al.\ 1996, Gibb et al.\ 2000).
Once the protostar has formed, its luminosity can heat the surrounding
grains to temperatures at which the ices evaporate back into the gas
phase, resulting in enhanced gas/solid ratios (e.g., van Dishoeck et
al.\ 1996, Dartois et al.\ 1998, Boonman et al.\ 2000). The
sublimation temperatures range from $\sim$20~K for pure CO ice to
$\sim$90~K for H$_2$O-rich ice under typical conditions. These freshly
evaporated molecules can then drive a rich and complex chemistry in
the gas (called the `hot core' phase) until the normal ion-molecule
chemistry takes over again after $\sim 10^5$ yr (Charnley et al.\
1992). The different solid-state and gas-phase species therefore serve
not only as physical diagnostics, but also as probes of the evolution
of the region.

\section{Our VLT-ISAAC program}

In 1999, we proposed a large VLT-UT1 ISAAC program to probe the origin
and evolution of ices in southern star-forming regions through a
spectroscopic survey of $30-40$ objects in the 2.7--5.1 $\mu$m L and
M-band atmospheric windows. The aim was to obtain high-quality spectra
($S/N>50$ on continuum) such that 3$\sigma$ limits of species with
abundances down to 2--4\% of H$_2$O ice --- the dominant ice component
--- can be obtained.  The program focuses on low- and
intermediate-mass southern YSOs, but covers a range of evolutionary
stages from background stars to T-Tauri stars with disks. Also,
several different environments (Vela, Chamaeleon, Ophiuchus, Corona
Australis, ...) will be probed.  The program was awarded 14 nights of
VLT time. However, due to technical problems with the long-wavelength
arm of ISAAC, the first observing session
did not take place until January 2001.

The January 2001 observations totalled 5 nights and were carried
out under mediocre conditions with high humidity. Nevertheless, good
low-resolution (LR, $R=600-800$) spectra were obtained for $\sim 15$
objects and medium-resolution (MR, $R=3000-5000$) spectra for $\sim 5$
objects, covering $\sim$one-third of our project.  With the new
1024$\times$1024 Aladdin array, the low-resolution spectra can be
obtained in a single spectral setting per atmospheric window, whereas
the medium resolution spectra require 6/3 settings to cover the entire
L/M band, respectively. Typical exposure times are $\sim$ 30 minutes
per atmospheric window in LR for a L$\approx 7$ mag object. Standard
stars were observed immediately before or after the YSOs, within
0.05--0.1 airmass. The spectra were obtained using both chopping (by
15$''$ along the slit) and nodding of the telescope. Daily arcs and
flat-fields were provided by the observatory staff. In the M-band, the
atmospheric CO lines were used for wavelength calibration. The data
were reduced using IDL routines developed in-house by E.\ Dartois and
W.F.\ Thi.

\begin{figure}[tbh]
\begin{center}
\includegraphics[width=.7\textwidth,angle=0]{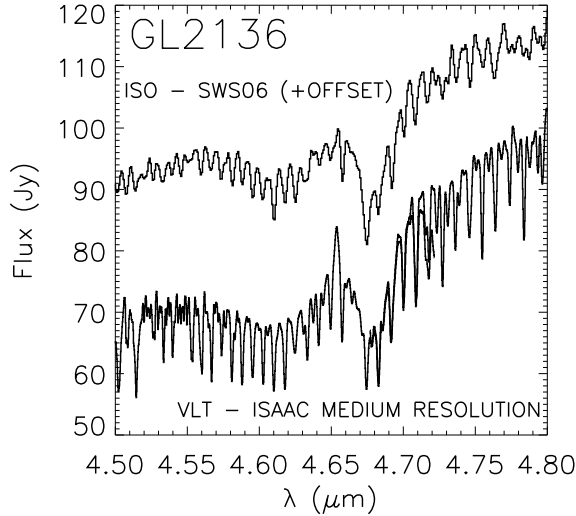}
\end{center}
\caption[]{VLT-ISAAC MR M-band $R\approx 2000$
spectrum of GL 2136 compared with the ISO-SWS
spectrum ($R\approx 1500$). 
The sharp features are due to gas-phase CO, and become
stronger in the higher-resolution VLT spectrum. Broad absorption features at 
4.62 $\mu$m due to solid OCN$^-$ and at 4.67
$\mu$m due to solid CO are seen as well. The H I 7--5 emission line 
at 4.6538 $\mu$m is 
stronger in the VLT spectrum due to the smaller slit.}
\label{fig1}
\end{figure}

Because the ice features are weak and broad and are superposed on a strong
continuum, many tests were carried out to check the reliability of the
spectra. Specifically, spectra were taken for objects also observed
with ISO or UKIRT, and features were checked on different nights at
the VLT and between the LR and MR modes on the VLT. In general, the
reproducibility of the features is excellent (see Fig.\ 1).

A particularly nice capability of ISAAC is the possibility to rotate the
slit on the sky to obtain spectra of more than one object
simultaneously. In several cases, more than one bright object at L or
M-band was discovered in the acquisition image within $\sim 15''$. The
slit was then rotated to include the additional object. An example is
shown in Fig.\ 2 for the high-mass source GL 961, where the E and W components
are separated by $\sim$5.5$''$.

Laboratory data such as those obtained in the Sackler Laboratory for
Astrophysics in Leiden and at the Astrochimie Exp\'erimentale
Laboratory in Paris play an essential role in the analysis of the
infrared spectra. First, they lead to definite identification of the
molecules and to quantitative estimates of their abundances. Second,
they provide an indication of the ice environment and ice
components. For example, a CO molecule embedded in an H$_2$O-rich
matrix (`polar ice') has a different spectral shape compared with that
in a CO-rich mantle (`apolar ice'). Such analyses have shown that the
ice mantles in the protostellar environment are not homogeneous, but
consist of several components. These phases may reflect differential
accretion of atomic H-rich versus H-poor gas and/or different degrees of
outgassing of the more volatile species (Schutte 1999). Third, the ice
spectra and the gas/solid ratios can provide clear evidence for heating of
the ices in the more evolved objects (Ehrenfreund et al.\ 1998,
Boogert et al.\ 2000, van Dishoeck et al.\ 1996).

\begin{figure}[tbh]
\begin{center}
\includegraphics[width=.8\textwidth,angle=90]{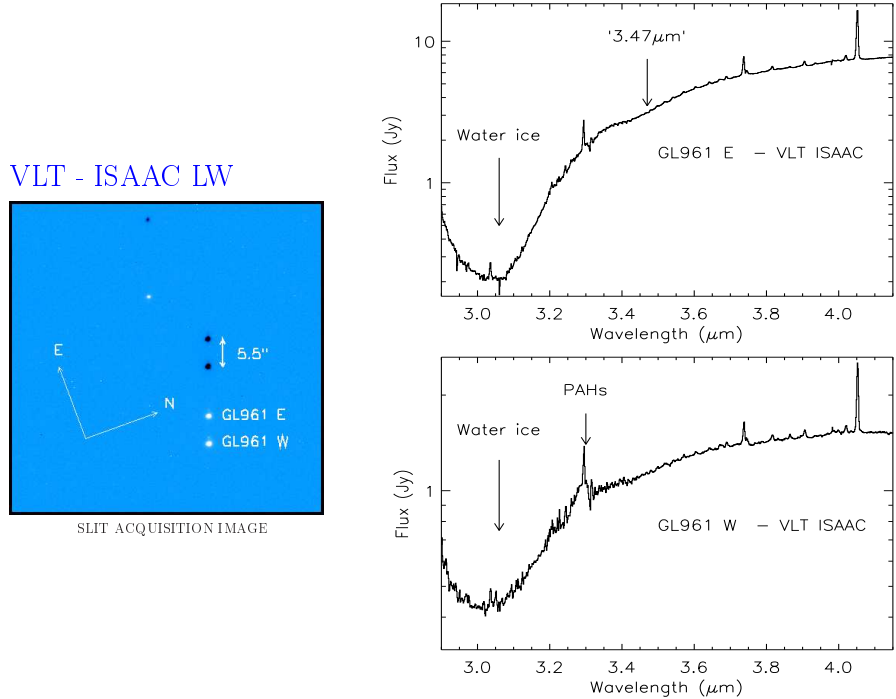}
\end{center}
\caption[]{VLT-ISAAC spectra obtained toward GL 961 E and W. The 3.3 $\mu$m
PAH feature may be affected by poor cancellation of an atmospheric feature.}
\label{fig2}
\end{figure}

\section{Initial results}

The prime targets in the January 2001 run were a set of low- and
intermediate mass YSOs in the Vela and Chamaeleon clouds, together
with a few well-known southern high-mass protostars. In the first
case, the power of the VLT+ISAAC is used to observe weaker and lower
luminosity sources than possible previously, whereas for the high-mass
sources the aim was to obtain higher $S/N$ and higher spectral
resolution data than provided by ISO or previous ground-based
observations.

The L- and M-band windows include the following main features:
3 $\mu$m (H$_2$O ice), 3.47 $\mu$m (unidentified), 3.54 $\mu$m
(CH$_3$OH ice), 4.08 $\mu$m (HDO ice), 4.62 $\mu$m (OCN$^-$ ice)
and 4.67 $\mu$m (CO gas and ice). The strong solid H$_2$O and CO bands
are detected in most objects, although with varying amounts.
Typical LR and MR spectra are shown in Figures 1--4. Because of the
strong atmospheric features at the relatively low altitude of Paranal,
the quality of the data around the solid OCS feature at 4.9 $\mu$m is
low.

Several of these features are excellent indicators of the thermal
history and energetic processing of the ices. Specifically, we can use
the following diagnostics of the physical conditions and evolution of
our sources: (i) the solid H$_2$O profile, where the peak position
gives an indication of the ice temperature; (ii) the solid CO profile,
where the shape indicates the `apolar' vs.\ `polar' ice fraction;
(iii) the CO/H$_2$O abundance ratio, with the more volatile CO
molecule having a lower ice abundance at high temperature; (iv) the
gas/solid CO ratio and the gas-phase CO excitation temperature; and
(v) the presence of the OCN$^-$ feature, which is thought to be a
tracer of energetic processing (ultraviolet irradiation or particle
bombardment) (Schutte \& Greenberg 1997, Demyk et al.\ 1998).  In the
following, a few specific initial results are presented.

\begin{figure}[tbh]
\begin{center}
\includegraphics[width=.7\textwidth]{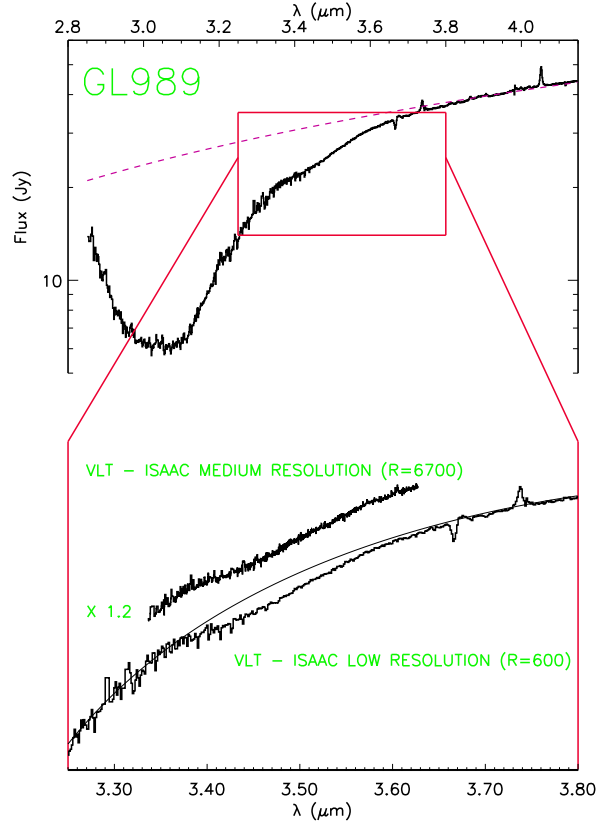}
\end{center}
\caption[]{VLT-ISAAC LR and MR L-band spectra obtained 
toward GL 989. Note the lack of substructure in the 3.47 $\mu$m feature
(Dartois et al., in prep).}
\label{fig3}
\end{figure}

\subsection{The 3.47$\mu$m feature}

The 3.47 $\mu$m feature, previously detected by Allamandola et al.\
(1992) and Brooke et al.\ (2000), is observed in several of our sources
at high $S/N$. The comparison of the LR and MR spectra for the bright
source GL 989 shows good agreement in the shape of the spectra (Figure
3). Also, no substructure is apparent at the higher spectral
resolution. The data are currently being compared with different
laboratory ice mixtures, in particular mixtures involving H$_2$O and
NH$_3$ (Dartois et al., in prep.).  Theoretical models predict that
NH$_3$ is an important component of interstellar ices formed by
hydrogenation of atomic N, but unfortunately the strongest NH$_3$
bands are blended with H$_2$O at 3 $\mu$m and with the silicate band
at 9.6 $\mu$m. A tentative detection of the 9.6 $\mu$m feature toward
one (northern) object has recently been claimed by Lacy et al.\
(1998), suggesting high NH$_3$ abundances up to 10\% with respect to H$_2$O
ice.  The analysis of our VLT data indicates lower NH$_3$ abundances.

\subsection{Vela sources}

LR L- and M-band spectra of 5 intermediate mass YSOs ($\sim 300-700$
L$_{\odot}$) have been obtained in the previously unexplored Vela
molecular cloud (Thi et al., in prep.). The objects were selected from
the list of `class I' objects of Liseau et al.\ (1992), based on their
spectral energy distribution. For two sources, an additional object was
found in the field within a few $''$, 
providing `off source' information on the
ices. MR spectra in the region
of the solid and gas-phase CO band have been taken as well for a few cases.

H$_2$O ice has been detected in all objects, and CO ice in half of the
objects.  The H$_2$O profile indicates that the bulk of the ice is
very cold.  The solid CO band toward IRAS 08375 --4109 is one of the
strongest and sharpest CO bands observed in any source: at LR, the
feature is unresolved and its true depth can only be obtained from the
MR spectrum (Figure 4). In spite of the low overall temperature, clear
differences in the solid CO/H$_2$O ice abundances are observed, which
correlate with the bolometric temperature of the source.

\begin{figure}[t]
\begin{center}
\includegraphics[width=.9\textwidth,angle=0]{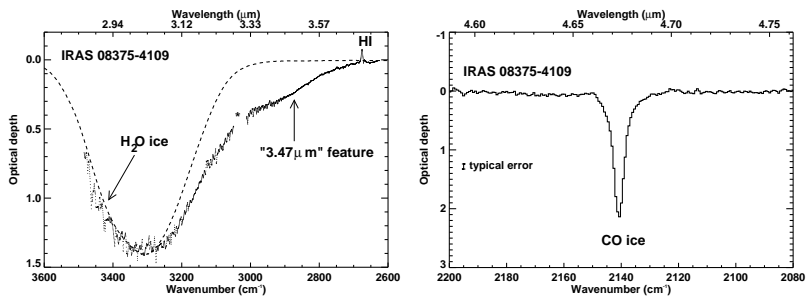}
\end{center}
\caption[]{VLT-ISAAC LR  L- and M-band spectra obtained 
toward IRAS 08375 --4109 in the Vela molecular cloud
(Thi et al., in prep).}
\label{fig4}
\end{figure}

Several of the sources show the 3.47 $\mu$m feature, but the presence of
the 3.54 $\mu$m band is less clear. This limits the solid CH$_3$OH
abundance to a few \% of that of H$_2$O ice, significantly less than
the 50\% found toward some massive protostars (Dartois et al.\
1999). The OCN$^-$ feature is detected in at least one high
temperature source.

\subsection{The circumstellar disk around L1489}

L1489 has been shown by Hogerheijde (2001, this volume) to be a transitional
object between the class I and II phases. It is surrounded by a large 2000~AU
radius rotating circumstellar disk, which must be on the verge of
shrinking to the $\sim 100$~AU size disks seen around T Tauri stars.
Thus, it provides an excellent opportunity to probe the chemical
composition of the gas and dust just when it is being incorporated into the
disk. The VLT spectra of L1489 show strong absorption by H$_2$O and CO ices,
but no evidence for CH$_3$OH and OCN$^-$ features, providing limits on their
abundances of a few~\% with respect to H$_2$O ice. \\

In summary, the initial data show that the VLT-ISAAC is a powerful
instrument to obtain high-quality 2.9--5 $\mu$m spectra of
low-luminosity embedded YSOs, and that such data can provide an
important step forward in our understanding of the physical and
chemical evolution of ices in low-mass young stellar objects and their
incorporation into new planetary systems.

%

\end{document}